\documentclass[twocolumn,showpacs,preprintnumbers,amsmath,amssymb,nofootinbib]{revtex4-1}
\usepackage{mathbbold}
\usepackage{amsmath}

\usepackage{graphicx}
\usepackage{dcolumn}
\usepackage{bm}

\begin{document}


\title{Covariance, correlation matrix and the multi-scale community structure of networks}
\author{Hua-Wei Shen}
\email{shenhuawei@software.ict.ac.cn}
\author{Xue-Qi Cheng}
\email{cxq@ict.ac.cn}
\author{Bin-Xing Fang}
\affiliation{Institute of Computing Technology, Chinese Academy of
Sciences, Beijing 100190, China}


\begin{abstract}
Empirical studies show that real world networks often exhibit
multiple scales of topological descriptions. However, it is still an
open problem how to identify the intrinsic multiple scales of
networks. In this article, we consider detecting the multi-scale
community structure of network from the perspective of dimension
reduction. According to this perspective, a covariance matrix of
network is defined to uncover the multi-scale community structure
through the translation and rotation transformations. It is proved
that the covariance matrix is the unbiased version of the well-known
modularity matrix. We then point out that the translation and
rotation transformations fail to deal with the heterogeneous
network, which is very common in nature and society. To address this
problem, a correlation matrix is proposed through introducing the
rescaling transformation into the covariance matrix. Extensive tests
on real world and artificial networks demonstrate that the
correlation matrix significantly outperforms the covariance matrix,
identically the modularity matrix, as regards identifying the
multi-scale community structure of network. This work provides a
novel perspective to the identification of community structure and
thus various dimension reduction methods might be used for the
identification of community structure. Through introducing the
correlation matrix, we further conclude that the rescaling
transformation is crucial to identify the multi-scale community
structure of network, as well as the translation and rotation
transformations.

\end{abstract}

\maketitle

\section{Introduction}
\label{introduction}

Many real world complex networks, including social
networks~\cite{Girvan02,Newman03a}, information
networks~\cite{Flake02, Cheng09}, and biological
networks~\cite{Girvan02,Guimera05}, are found to divide naturally
into communities, known as groups of nodes with a
higher-than-average density of edges connecting them. Communities
are of interest because they often correspond to functional units
such as collections of web pages on a single topic. The
identification of community structure has attracted much attention
in various scientific fields. Many methods have been proposed and
applied successfully to some specific complex
networks~\cite{Newman04,Clauset04,Radicchi04,Palla05,Duch05,Newman06a,
Newman06b,Shen09a,Shen09b,Bagrow08,Carmi08}. For reviews, the reader
can refer to~\cite{Fortunato10}.

The community structure of real world networks often exhibit
multiple scales~\cite{Fortunato07,Lambiotte08,Arenas08,Ronhovde09}.
However, the multi-scale community structure cannot be uncovered
directly through traditional methods. Arenas \emph{et
al}~\cite{Arenas06} pointed out that the synchronization process
reveals topological scales of networks and that the spectrum of the
Laplacian matrix can be used to identify such topological scales.
Cheng and Shen~\cite{Cheng10} proposed the network conductance to
identify the multiple topological scales through investigating the
diffusion process taking place on networks. Recently,
in~\cite{Mucha10}, a general framework is proposed for the detection
of community structure in multi-scale networks.

Generally, the straightforward description of network topology is a
high-dimensional but redundant description, where each node is taken
as one dimension of the network and the edges characterize the
relationship between these dimensions. The identification of
community structure can be viewed as finding the most significant
reduced dimensions which capture the main characteristics of network
topology~\cite{Arenas10}. Different significance levels for such
dimensions correspond to the community structure with different
topological scales.

In this article, we first apply the \emph{principal component
analysis (PCA)} to characterize the community structure. We show
that the covariance matrix behind the PCA is the unbiased version of
the modularity matrix~\cite{Newman06b}, which underlies the
well-known benefit function modularity for community detection. From
the perspective of dimension reduction, however, the covariance
matrix only takes into account the translation and rotation
transformations. These two transformations fail to deal with the
heterogeneous distribution of node degree and community size. To
address this problem, a correlation matrix is proposed through
introducing the rescaling transformation into the covariance matrix.
Theoretical analysis indicates that all these three transformations
are crucial to identify the multi-scale community structure.
Finally, the effectiveness of the correlation matrix is demonstrated
through the comparison with the covariance matrix or the modularity
matrix on real world networks and artificial benchmark networks.
Extensive tests demonstrate that the correlation matrix is very
effective at uncovering the multi-scale community structure of
network and it significantly outperforms the modularity matrix or
the covariance matrix, especially when the distribution of node
degree are heavily heterogenous. The dimension reduction perspective
opens the door to utilizing various dimension reduction techniques
for the identification of community structure.

\section{Covariance matrix of network}
\label{sec2}

A directed network is often described by its adjacency matrix $A$
whose element $A_{ij}$ being $1$ if there exists an edge pointing to
node $j$ from node $i$, and $0$ otherwise. The node $i$ is called
the tail of the edge and the node $j$ is called the head of the
edge. The out-degree of a node is defined as
$k_i^{out}=\sum_j{A_{ij}}$ and the in-degree is
$k_j^{in}=\sum_i{A_{ij}}$. We suppose that the network has $n$ nodes
and $m$ edges with $m=\sum_i{k_i^{out}}=\sum_j{k_j^{in}}$. For an
undirected network, it can be transformed into a directed one by
replacing each edge with two oppositely directed edges. Note that,
for a self-loop edge, only one directed edge is used to replace the
original undirected one.

Another representation for network is given by two node-edge
incidence matrices with the size $n$ by $m$, which are respectively
defined as
\begin{equation}
X_{il} = \begin{cases}
           \enspace 1 & \text{if the node $i$ is the tail of the edge $l$,} \\
           \enspace 0 & \text{otherwise,}
         \end{cases}
\label{eq1}
\end{equation}
and
\begin{equation}
Y_{il} = \begin{cases}
           \enspace 1 & \text{if the node $i$ is the head of the edge $l$,} \\
           \enspace 0 & \text{otherwise.}
         \end{cases}
\label{eq2}
\end{equation}
Note that the rows of $X$ (or $Y$) are mutually orthogonal and that
the columns each sum to unity.

As to this representation, $n$ nodes correspond to $n$ dimensions
with the $i$th dimension being denoted by $e_{i}$, whose $i$th
element is $1$ and other elements are $0$. The columns of $X$ or $Y$
can be taken as $n$-dimensional data points distributed in the space
spanned by the $n$ dimensions. The mean of these data points in $X$
is denoted by $x=\frac{1}{m}\sum_j{X_{*j}}$, where $X_{*j}$ is the
$j$th column of $X$. Similarly, we give the mean of the data points
in $Y$ as $y=\frac{1}{m}\sum_j{Y_{*j}}$. According to
Eq.~(\ref{eq1}) and Eq.~(\ref{eq2}), we have $x=\big(k_1^{out},
k_{2}^{out},\cdots,k_{n}^{out})^T/m$ and $y=\big(k_1^{in},
k_{2}^{in},\cdots,k_{n}^{in})^T/m$. Now we subtract off the mean $x$
from each column of $X$ and the mean $y$ from each column of $Y$.
Such an operation is known as the \emph{translation} transformation
and makes the data points mean zero. The resulting matrices are
denoted by $\tilde{X}=X-x\mathbbold{1}^T$ and
$\tilde{Y}=Y-y\mathbbold{1}^T$, where $\mathbbold{1}$ is the
$m$-dimensional vector with all its elements being $1$. With
$\tilde{X}$ and $\tilde{Y}$, the covariance between the $i$th row of
$X$ and the $j$th row of $Y$ can be calculated by
$\tilde{X}_{i*}\cdot(\tilde{Y}_{j*})^{T}/(m-1)$. Here, the
denominator $m-1$ is used instead of $m$ to make the covariance be
unbiased. In this way, all these covariances between the rows of $X$
and the rows of $Y$ form an $n\times n$ matrix
\begin{equation}
C=\frac{1}{m-1}\tilde{X}\tilde{Y}^T.\label{eq3}
\end{equation}
This matrix is referred to as the \emph{covariance matrix of
network}. Its elements are
\begin{equation}
C_{ij} =
\frac{1}{m-1}\left(A_{ij}-\frac{k_{i}^{out}k_{j}^{in}}{m}\right).
\label{eq4}
\end{equation}
Note that the covariance matrix can be easily extended to weighted
networks if we consider each weighted edge between two nodes as
multiple unweighted edges connecting them.

To our surprise, the covariance matrix is identical to the
modularity matrix except that the denominator in the first term is
$m-1$ in the covariance matrix while it is $m$ in the modularity
matrix. In statistics, when calculating the empirical variance from
sample data points rather than the distribution itself, $m-1$ is
used instead of $m$ to make the empirical variance be unbiased. Thus
the covariance matrix could be taken as the unbiased version of the
modularity matrix.

\section{Spectrum of the covariance matrix and community structure}
\label{sec3}

Matrix spectral analysis provides an important technique for network
division and the identification of community structure. For example,
as to the Laplacian matrix, the Fielder's
eigenvector~\cite{Fiedler73} is well studied and widely used for
two-way network partition. Newman proposed to find the community
structure of network using the eigenvectors of the modularity
matrix~\cite{Newman06a}. This section aims to illustrate what is
told about the community structure by the covariance matrix from the
perspective of dimension reduction.

It is well known that the traditional covariance matrix plays an
important role in PCA. By analyzing its eigenvalues and
eigenvectors, the most significant directions (or dimensions) in
terms of the variance are identified as principal components. The
lesser significant ones are discarded to reduce the number of
dimensions and to alleviate redundance without losing too much
relevant information. In this article, the idea similar to PCA is
adopted to analyze the role of the covariance matrix of network at
uncovering the community structure of network.

Intuitively, as to networks with community structure, the tail nodes
of edges are expected to be positively correlated with the head
nodes of edges~\cite{Newman02b,Newman03b}. According to the
definition of covariance, the more two variables correlate with each
other, the larger the covariance between them. Now the task is to
find a direction along which the covariance between $\tilde{X}$ and
$\tilde{Y}$ reaches the maximum. Without loss of generality, we use
a normalized vector $v$ to denote such a direction. We write $v$ as
a linear combination of the normalized eigenvectors $v_i$ of the
covariance matrix $C$, i.e., $v=\sum_{i}^{n}{a_{i}v_{i}}$, where the
coefficients $a_i=v_{i}^{T}v$. Since that $v$ is a normalized
vector, we have $v^{T}v=1$ and this implies that
\begin{equation}
\sum_{i=1}^{n}{a_{i}^{2}}=1. \label{eq5}
\end{equation}

Due to that $\tilde{X}$ and $\tilde{Y}$ have zero means, the
covariance between them along the direction $v$ can be calculated by
\begin{eqnarray}
V&=&\frac{1}{m-1}(v^{T}\tilde{X})(v^{T}\tilde{Y})^{T}=\frac{1}{m-1}(v^{T}\tilde{X})(\tilde{Y}^{T}v) \nonumber \\
 &=&v^{T}Cv=(\sum_{i}^{n}{a_{i}v_{i}^{T}})C(\sum_{j}^{n}{a_{j}v_{j}})
 \nonumber \\
 &=&\sum_{ij}{a_{i}a_{j}\lambda_{j}\delta_{ij}}=\sum_{i}{a_{i}^{2}\lambda_{i}},\label{eq6}
\end{eqnarray}
where $\lambda_{i}$ is the eigenvalue of $C$ corresponding to the
eigenvector $v_{i}$ and we have made use of
$v_{i}^{T}v_{j}=\delta_{ij}$. Without loss of generality, we assume
that the eigenvalues are labeled in decreasing order
$\lambda_{1}\geq\lambda_{2}\geq\cdots\geq\lambda_{n}$\footnote{In
the case that the eigenvalues are complex numbers, the eigenvalues
are ranked in the decreasing order of the real part of these complex
numbers.}. The task of maximizing $V$ can then be equated with the
task of choosing the nonnegative quantities $a_{i}^{2}$ so as to
place as much as possible of the weight in the sum (\ref{eq6}) in
the terms corresponding to the most positive eigenvalues, and as
little as possible in the terms corresponding to the most negative
ones, while respecting the normalization constraint in
Eq.~(\ref{eq5}).

Obviously, $V$ reaches maximum when we set $a_{1}^2=1$ and
$a_{i}^2=0 (i\neq{1})$, i.e., the desired direction $v$ is parallel
to the eigenvector $v_{1}$. Then, we turn to the next direction
along which the covariance is maximized with the constraint that it
is orthogonal to the obtained direction $v_1$. According to
Eq.~(\ref{eq5}) and Eq.~(\ref{eq6}), such a direction is parallel to
the eigenvector $v_2$. In the similar way, it can be easily shown
that all the eigenvectors of the covariance matrix $C$ form a set of
orthogonal bases for the $n$-dimensional space, with these
dimensions being ranked in terms of the covariance between
$\tilde{X}$ and $\tilde{Y}$ along them.

In the following, we will show that the community structure of
network is reflected by eigenvectors of the covariance matrix of
network. Specifically, the eigenvectors corresponding to positive
eigenvalues make positive contribution to reflect the community
structure of network, and the negative ones provide information for
the anticommunity structure of network. This phenomenon has been
noticed by Newman when studying the spectrum of the modularity
matrix~\cite{Newman06b}. Different from Newman's work, we illustrate
the relation between the covariance matrix and the community
structure from the perspective of dimension reduction.

We use $U$ to denote the eigenvector matrix with its columns being
the normalized eigenvectors of the covariance matrix $C$ ranked in
the descending order of the corresponding eigenvalues, which are
placed on the diagonal of the diagonal matrix $D$. Then we have
\begin{eqnarray}
D &=& U^{T}CU \nonumber \\
  &=& \frac{1}{m-1}U^{T}(\tilde{X}\tilde{Y}^{T})U \nonumber \\
  &=& \frac{1}{m-1}(U^{T}\tilde{X})(U^{T}\tilde{Y})^{T}.\label{eq7}
\end{eqnarray}

Multiplying a vector by $U^{T}$ from the left-hand side means
reexpressing it with respect to the new orthogonal bases comprising
of the columns of $U$. Thus, $\hat{X}=U^{T}\tilde{X}$ and
$\hat{Y}=U^{T}\tilde{Y}$ are respectively the new coordinates of the
original sample data $\tilde{X}$ and $\tilde{Y}$ with respect to the
new bases. Note that $\hat{X}$ and $\hat{Y}$ are respectively the
results of rotating $\tilde{X}$ and $\tilde{Y}$ around the
coordinate origin. Thus, $\hat{X}$ and $\hat{Y}$ both have zero mean
and accordingly the matrix $D$ is the covariance matrix for
$\hat{X}$ and $\hat{Y}$. All the off-diagonal elements of $D$ are
zeroes, indicating that $\hat{X}$ only correlates to $\hat{Y}$ along
the same dimension and they are linearly independent along different
dimensions. If $\hat{X}$ and $\hat{Y}$ are positively correlated
along a certain dimension, the corresponding diagonal element of $D$
is positive, otherwise negative. Furthermore, the magnitude of the
eigenvalues provides an indicative index for the significance of the
dimensions to reflect the community structure of the underlying
network. Hereafter, we will neglect the constant multiplier
$1/(m-1)$ in the covariance matrix without losing any information
for the detection of the community structure.

\begin{figure}
\includegraphics[width=0.48\textwidth]{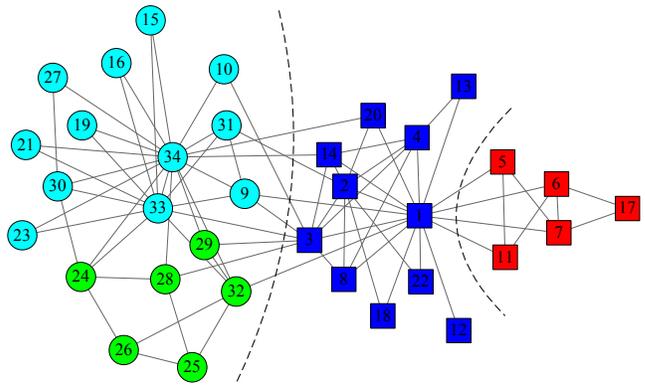}
\caption{\label{fig1} (Color online) The network of the karate club
studied by Zachary~\cite{Zachary77}. The real social fission of this
network is represented by two different shapes, circle and square.
Different colors depict the communities at the other scale. The
dashed-curve gives another alternative partition of network.}
\end{figure}

\begin{table}
\caption{\label{tab1}Eigenvectors corresponding to the most positive
and negative eigenvalues of the covariance matrix of Zachary's club
network}
\begin{ruledtabular}
\begin{tabular}{rrr|rrr}
node & positive & negative & node & positive & negative \\
\hline
1  & -0.3875 &  0.4576 & 18 & -0.1320 & -0.0720 \\
2  & -0.2696 &  0.1957 & 19 &  0.1394 & -0.1259 \\
3  & -0.1319 &  0.1371 & 20 & -0.0576 & -0.1605 \\
4  & -0.2535 &  0.0376 & 21 &  0.1394 & -0.1259 \\
5  & -0.1340 & -0.0127 & 22 & -0.1320 & -0.0720 \\
6  & -0.1457 &  0.0018 & 23 &  0.1394 & -0.1259 \\
7  & -0.1457 &  0.0018 & 24 &  0.2167 & -0.0540 \\
8  & -0.2094 & -0.0585 & 25 &  0.0563 &  0.0862 \\
9  &  0.0545 & -0.1490 & 26 &  0.0754 &  0.0867 \\
10 &  0.0479 & -0.0906 & 27 &  0.1158 & -0.0551 \\
11 & -0.1340 & -0.0127 & 28 &  0.1028 & -0.0515 \\
12 & -0.0778 & -0.0594 & 29 &  0.0683 & -0.0429 \\
13 & -0.1287 & -0.0438 & 30 &  0.2063 & -0.0616 \\
14 & -0.1350 & -0.1469 & 31 &  0.0963 & -0.0894 \\
15 &  0.1394 & -0.1259 & 32 &  0.1019 & -0.1414 \\
16 &  0.1394 & -0.1259 & 33 &  0.3239 &  0.3346 \\
17 & -0.0585 &  0.0442 & 34 &  0.3698 &  0.6200
\end{tabular}
\end{ruledtabular}
\end{table}

To illustrate what is provided about the community structure of
network by the covariance matrix, as an example, we analyze the
covariance matrix of the Zachary's karate club
network~\cite{Lusseau03}, which is widely used as a benchmark for
the methods of community detection. The network and its community
structure are depicted in Fig.~\ref{fig1}. Table~\ref{tab1} shows
the eigenvectors corresponding to the most positive and negative
eigenvalues of the covariance matrix. We can see that the real
fission of the club network is exactly revealed by the signs of the
components in the eigenvector corresponding to the most positive
eigenvalue. As to the eigenvector corresponding to the most negative
one, its components divide the nodes into two groups with many
connections lying between them, reflecting the so-called
anticommunity structure.

Besides the eigenvector corresponding to the most positive
eigenvalue, other eigenvectors corresponding to positive eigenvalues
can also be utilized to reveal the community structure of network.
Specifically, the components of the first $p$ eigenvectors are used
to represent the nodes of network into $p$-dimensional vectors and
the community structure is then uncovered through clustering the
node vectors using, for example, the $k$-means clustering method. We
have known that the original description of network topology, i.e.,
Eq.~(\ref{eq1}) and Eq.~(\ref{eq2}), actually uses a set of standard
bases. With respect to this set of bases, the $i$th node of network
is represented with the coordinate vector $e_{i}$ whose elements are
both $0$ but the $i$th one being $1$. The columns of the normalized
eigenvector matrix $U$ constitute another set of orthogonal bases,
which can be obtained from the standard bases through the
\emph{rotation} transformation, i.e., multiplying the standard bases
with $U^T$ from the left-hand side. With respect to this new bases,
the coordinate vector of the $i$th node can be represented by
$U^{T}e_{i}$, which is the projection of the original coordinate
vector on the new set of bases. Mathematically, the $i$th projected
node vector $r_{i}$ can be denoted by
\begin{equation}
[r_{i}]_{_{j}}=U_{ij}.\nonumber
\end{equation}
Note that the covariances along different new axis direction
$U_{*j}$ between the data $\tilde{X}$ and $\tilde{Y}$ vary,
characterized by
\begin{equation}
C=UDU^T=(UD^{1/2})(UD^{1/2})^T.\nonumber
\end{equation}
Thus, for the purpose of clustering the node vectors, it is more
appropriate to represent the $i$th node with a node vector $r_{i}$
with its $j$th component being
\begin{equation}
[r_{i}]_{_{j}}=\sqrt{\lambda_{j}}U_{ij}.\label{eq8}
\end{equation}
For negative eigenvalues, the corresponding components of the node
vector are complex numbers.

Using Eq.~(\ref{eq8}), we can partition the nodes of network into
communities through clustering the node vectors $r_{i}$ using the
$k$-means clustering method. However, only the eigenvectors
corresponding to positive eigenvalues make positive contribution to
the community structure. Furthermore, the eigenvectors corresponding
to the small positive eigenvalues are less significant to uncover
the community structure of network. This poses a challenging
problem, i.e., the choice of the number of eigenvectors.

Intuitively, as for the goal of identifying the community structure,
all the eigenvectors corresponding to positive eigenvalues provide
certain relevant information. This means that the number of positive
eigenvalues gives an upper bound on the number of eigenvectors used.
As a kind of mesoscopic structure of network, however, the community
structure provides a coarse-grained description of the network
topology. Only the most significant structural features are
maintained and the less ones are neglected. Now the tricky problem
is how to determine the significant ones. In this article, this
problem is equivalent to the choice of the significant eigenvectors.
We have known that the magnitudes of eigenvalues characterize the
covariance between the data $\tilde{X}$ and $\tilde{Y}$ along the
direction of the corresponding eigenvectors. Thus we can choose the
number of significant eigenvectors according to the magnitude of
eigenvalues. Intuitively, the eigenvectors corresponding to larger
positive eigenvalues are desired. Furthermore, a large eigengap,
i.e., interval between two successive eigenvalues, provides an
effective indicator to determine the appropriate number of
significant eigenvectors. Similar methods have been adopted in other
contexts taking the advantage of the eigengap of many other types of
matrix, including the adjacency matrix~\cite{Chauhan09}, the
transition matrix~\cite{Eriksen03,Capocci05}, the Laplacian
matrix~\cite{Arenas06}, the modularity matrix~\cite{Newman06b} and
the normalized Laplacian matrix~\cite{Cheng10}. More importantly,
the choice of eigenvectors with different significance levels
determines the different scale of community structure. The existence
of a significant scale is indicated by the occurrence of a larger
eigengap in the spectrum of the covariance matrix.

Another important problem is the choice of the number of
communities. According to Eq.~(\ref{eq8}), we know that each node of
network can be represented by a $p$-dimensional node vector through
projecting its standard coordinate vector on the directions of the
$p$ significant eigenvectors. Then the identification of community
structure becomes a problem of partitioning the node vectors into
groups. According to~\cite{Newman06b}, the number of communities is
one greater than the number of orthogonal dimensions to represent
these vectors, i.e., $p+1$ is the appropriate number of community
when the top $p$ eigenvector is adopted to obtain the projected node
vectors.

\begin{figure}
\includegraphics[width=0.48\textwidth]{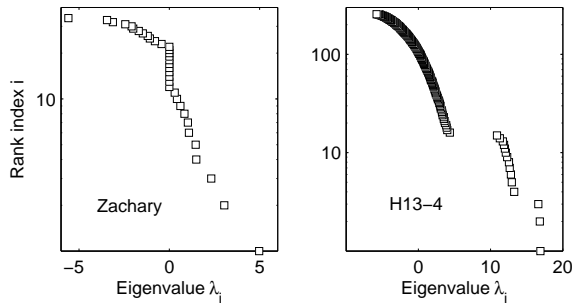}
\caption{\label{fig2} The spectrum of the covariance matrix. Left
panel: the Zachary's karate club network; Right panel: the H13-4
network.}
\end{figure}

Taking the Zachary's karate club network as example again, we
illustrate the effectiveness of the eigengap at determining the
number of significant eigenvectors and accordingly the number of
communities. As shown in Fig.~\ref{fig2} (left panel), the largest
eigengap (only the ones between positive eigenvalues are considered)
occurs between the first and second largest eigenvalues. It
indicates that it is appropriate to choose only the first
eigenvector and the number of communities is $2$. Furthermore, the
resulting two communities exactly reflect the real split of the
network in Fig.~\ref{fig1}. In addition, besides the largest
eigengap, two other relative larger eigengaps can be observed, one
being between the second and the third eigenvalues, and the other
being between the third and the forth eigenvalues. The resulting
partition according to these two eigengaps are also depicted in
Fig.~\ref{fig1}, one dividing the network into three communities
separated using dashed curves, and the other dividing the network
into four communities differentiated by colors. These two partitions
are often the results of many community detection methods. Although
they are not identical to the real split of the network, they reveal
certain relevant topological feature of the network at alternative
scales.

Actually, as to a network with multi-scale community structure, each
scale corresponds to a large eigengap in the spectrum of the
covariance matrix. Thus we can identify the multi-scale community
structure using the top eigenvectors indicated by these eigengaps.
As an example, we illustrate the identification of the multi-scale
community structure of the H13-4 network, which is constructed
according to~\cite{Arenas06}. The network has two predefined
hierarchical levels. The first hierarchical level consists of $4$
groups of $64$ nodes and the second hierarchical level consists of
$16$ groups of $16$ nodes. On average, each node has $13$ edges
connecting to the nodes in the same group at the second hierarchical
level and has $4$ edges connecting to the nodes in the same group at
the first hierarchical level. This explains the name of such kind of
networks. In addition, the average degree of each node is $18$.
According to the construction rules of the H13-4 network, the two
hierarchical levels constitute the different topological description
of the community structure of the H13-4 network at different scales.
As shown in Fig.~\ref{fig2} (right panel), the largest eigengap
occurs between the $15$th and $16$th largest eigenvalues, indicating
that the community structure at the most significant scale
corresponds to the partition dividing the nodes into $16$ groups.
Actually, the resulting communities are exactly the predefined $16$
groups of $16$ nodes in the second hierarchical level. In addition,
the second largest eigengap occurs between the third eigenvalue and
the forth one, indicating that the second significant topological
scale corresponds to the partition dividing the network nodes into
$4$ groups. Again the resulting communities are exactly the
predefined $4$ groups of $64$ nodes in the first hierarchical level.

\section{Finding the multi-scale community structure using the correlation matrix}
\label{sec4}

The previous section shows that the spectrum of the covariance
matrix provides an promising way to reveal the multi-scale community
structure of network. Note that, as to the example H13-4 network,
the nodes have approximately the same degree and the communities at
a specific scale are of the same size. However, the real world
networks usually have heterogenous distributions of node degree and
community size. Thus it will be more convincing to test the
covariance matrix on networks with heterogenous distributions of
node degree and community size.

\begin{figure}
\includegraphics[width=0.4\textwidth]{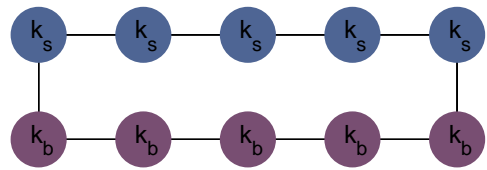}
\caption{\label{fig3}(Color online) The clique circle network as a
schematic example. Each circle corresponds to a clique, whose size
is marked by its label. The cliques labeled with $k_s$ are smaller
cliques with the size $s$, while the cliques labeled with $k_b$ are
bigger cliques with the size $b$. In this article, $s=10$ and
$b=20$.}
\end{figure}

\begin{figure}
\includegraphics[width=0.48\textwidth]{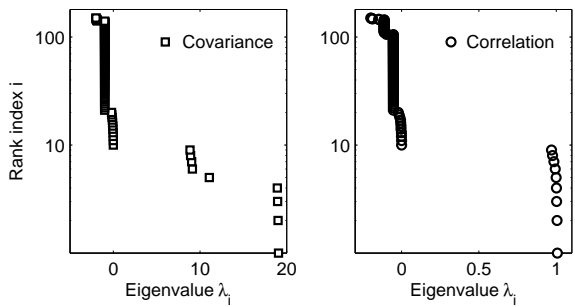}
\caption{\label{fig4} The spectrum of the covariance matrix (left
panel) and the correlation matrix (right panel) corresponding to the
clique circle network depicted in Fig.~\ref{fig3}. The horizontal
axis shows the eigenvalues of matrix and the vertical axis
represents the rank index of eigenvalues in descending order.}
\end{figure}

Before we give such a test in the subsequent section, using a
schematic network, we first illustrate the ineffectiveness of the
covariance matrix to deal with the heterogeneous distributions of
node degree and community size. The schematic network is often
called the clique circle network as depicted in Fig.~\ref{fig3}.
Generally speaking, the intrinsic community structure corresponds to
the partition where each clique is taken as a community, that is,
only one intrinsic scale exists in this network. However, as shown
in Fig.~\ref{fig4} (left panel), two scales are observed when we
investigate the community structure of this network using the
spectrum of the covariance matrix. One scale corresponds the
intrinsic scale of the network, and the other corresponds to
dividing the network nodes into $5$ groups, which is not desired.

To address this problem, we reconsider the formulation of the
covariance matrix in Section~\ref{sec2} and the spectral analysis on
the covariance matrix in Section~\ref{sec3}. When formulating the
covariance matrix from the data defined in Eq.~(\ref{eq1}) and
Eq.~(\ref{eq2}), the mean is subtracted off to place the data around
the coordinate origin. This is the translation transformation. When
using the eigenvectors of the covariance matrix as the new
orthogonal bases instead of the standard bases, the rotation
transformation is adopted. These two transformations, however, both
do not take into account the difference amongst the variances of the
original dimensions each corresponding to a node. Thus the spectrum
of the covariance matrix fails to deal with the heterogeneous
distribution of node degree and community size. Fortunately, the
possible remedy is also right here, i.e., the rescaling
transformation. Through introducing the rescaling transformation
into the covariance matrix, we obtain a correlation matrix, which
can be formulated as
\begin{equation}
R = (\Sigma_{X})^{-1}C(\Sigma_{Y})^{-1},\label{eq9}
\end{equation}
where $\Sigma_{X}$ is a diagonal matrix with its diagonal elements
$(\Sigma_{X})_{ii}=\sqrt{k_{i}^{out}(1-k_{i}^{out}/m)/(m-1)}$ being
the empirical variance of the data $X$ along the $i$th standard axis
direction and $\Sigma_{Y}$ is a diagonal matrix with its diagonal
elements $(\Sigma_{Y})_{jj}=\sqrt{k_{j}^{in}(1-k_{j}^{in}/m)/(m-1)}$
being the empirical variance of the data $Y$ along the $j$th
standard axis direction. Specifically, the elements of $R$ are
defined as
\begin{equation}
R_{ij}=
\frac{A_{ij}-\frac{k_{i}^{out}k_{j}^{in}}{m}}{\sqrt{k_{i}^{out}(1-k_{i}^{out}/m)}\sqrt{k_{j}^{in}(1-k_{j}^{in}/m)}}.
\label{eq10}
\end{equation}

Compared with the covariance matrix, the correlation matrix has two
advantages at identifying the multi-scale community structure. On
one hand, the correlation matrix can well deal with the
heterogeneous distribution of node degree and community size. As
shown in Fig.~\ref{fig4} (right panel), the intrinsic scale of the
clique circle network is correctly revealed by the spectrum of the
correlation matrix. On the other hand, the eigenvalues of the
correlation matrix themselves can provide intuitive judgements for
the cohesiveness within communities and the looseness of connections
among different communities. As shown in Fig.~\ref{fig4}, the
eigenvalues lying the greater side of the largest eigengap all
approach $1$. This indicates that the intrinsic communities are very
cohesive. Meanwhile, other eigenvalues are very small, indicating
the loose connectivity among these communities. Specifically,
through introducing the rescaling transformation into the covariance
matrix, the eigenvalues of the correlation matrix are rescaled.
Thus, besides the eigengaps among successive eigenvalues, the
eigenvalues themselves also provide indicative information of the
community structure of network. This is especially important for the
networks without significant topological scale. For these networks,
the eigengaps of the covariance matrix or the correlation matrix
both fail to provide obvious evidence for the number of intrinsic
communities. Without the rescaling transformation, the magnitude of
the covariance matrix is influenced by the network size and the
distribution of node degree. Thus the magnitude itself cannot
provide useful information to determine the cohesiveness of the
detected communities. As to the correlation matrix, however, the
magnitude of the eigenvalues of the correlation matrix has been
rescaled and thus can provide intuitive knowledge about the
cohesiveness of communities and then can help us choose the desired
scale with respect to specific application demands. More
importantly, the eigenvalues of the correlation matrix can be
compared among different networks due to that they are rescaled and
thus not influenced by the network size. In the subsequent section,
the effectiveness of the correlation matrix will be further
demonstrated through extensive tests on artificial benchmark
networks and real world networks.

Now we clarify the implication of the correlation matrix to the
widely-used benefit function modularity for network partition. As
shown in~\cite{Newman06a,Newman06b}, the modularity matrix (the
biased version of the covariance matrix used in this article)
underlies the benefit function modularity and the optimization of
the modularity can be carried out using the eigenvectors of the
modularity matrix. In this article, we propose to uncover the
community structure using the eigenvectors of the correlation
matrix, which is obtained through introducing the rescaling
transformation into the covariance matrix. In this way, the benefit
function for network partition actually changes. Specifically, the
correlation matrix provides us another benefit function for network
partition. This benefit function can be formulated as
\begin{equation}
F = \sum_{c}{\sum_{ij}{\delta_{ic}\delta_{jc}R_{ij}}},\label{eq11}
\end{equation}
where $c$ denotes a community, and $\delta_{ic}=1$ if node $i$
belongs to the community $c$ and $0$ otherwise. On one hand, the
benefit function $F$ can be directly optimized to uncover the
community structure similar to the role of the modularity. On the
other hand, the rationale behind $F$ and the modularity are somewhat
different. It is well known that the success of the modularity is
partly from its considering the difference between the actual weight
of the edge connecting node $i$ and $j$ and the expected weight of
the edge by chance characterized usually using a null model, e.g.
the configuration model. In this way, the modularity takes into
account the effects from the heterogeneous node degrees. However,
the difference of the variance associated with each node dimension
is not taken into account in the definition of modularity. Thus the
modularity may be blind to smaller communities when the network has
heterogeneous distribution of node degree and community size. This
problem roots in the heterogeneity of network rather than the
existence of an intrinsic scale (usually depends on the size of
network) causing the resolution limit problem~\cite{Fortunato07} of
modularity in the definition of modularity. This
heterogeneity-caused problem can be well dealt with by the new
benefit function $F$ since that the underlying correlation matrix
takes into account the rescaling transformation besides the
translation transformation and the rotation transformation utilized
in the the definition of modularity. In summary, it is inadequate to
only consider the heterogeneous node degree through using a
reference null model as done by the modularity. A rescaling
transformation is required to combat such a drawback facing the
modularity.

\section{Experimental results}
\label{sec5}

In this section, we empirically demonstrate the effectiveness of the
multi-scale community detection methods based on the spectrum of the
covariance matrix and the correlation matrix through tests on the
artificial benchmark networks. In addition, we apply the correlation
matrix method on a variety of real world networks.

\subsection{Tests on artificial benchmark networks}

We utilize the benchmark proposed by Lancichinetti \textit{et al}
in~\cite{Lancichinetti08}. This benchmark provides networks with
heterogeneous distributions of node degree and community size. Thus
it poses a much more severe test to community detection algorithms
than standard benchmarks. Many parameters are used to control the
generated networks in this benchmark: the number of nodes $N$, the
average node degree $\langle k\rangle$, the maximum node degree
max$\rule[-1pt]{0.15cm}{0.3pt}k$, the mixing ratio $\mu$, the
exponent $\gamma$ of the power law distribution of node degree, the
exponent $\beta$ of the power law distribution of community size,
the minimum community size min$\rule[-1pt]{0.15cm}{0.3pt}c$, and the
maximum community size max$\rule[-1pt]{0.15cm}{0.3pt}c$. In our
tests, we use the default parameter configuration where $N=1000$,
$\langle k\rangle=15$, max$\rule[-1pt]{0.15cm}{0.3pt}k=50$,
min$\rule[-1pt]{0.15cm}{0.3pt}c=20$, and
max$\rule[-1pt]{0.15cm}{0.3pt}c=50$. To test the influence from the
distribution of node degree and community size, we adopt four
parameter configurations for $\gamma$ and $\beta$, respectively
being $(\gamma,\beta)=(2,1)$, $(\gamma,\beta)=(2,2)$,
$(\gamma,\beta)=(3,1)$ and $(\gamma,\beta)=(3,2)$ . By tuning the
mixing ratio $\mu$, we test the effectiveness of our method on the
networks with different fuzziness of communities. The larger the
parameter $\mu$, the fuzzier the community structure of the
generated network. In addition, we adopt the normalized mutual
information (NMI)~\cite{Danon2005} to compare the partition found by
community detection methods with the answer partition. The larger
the NMI, the more effective the tested method.

\begin{figure}
\includegraphics[width=0.48\textwidth]{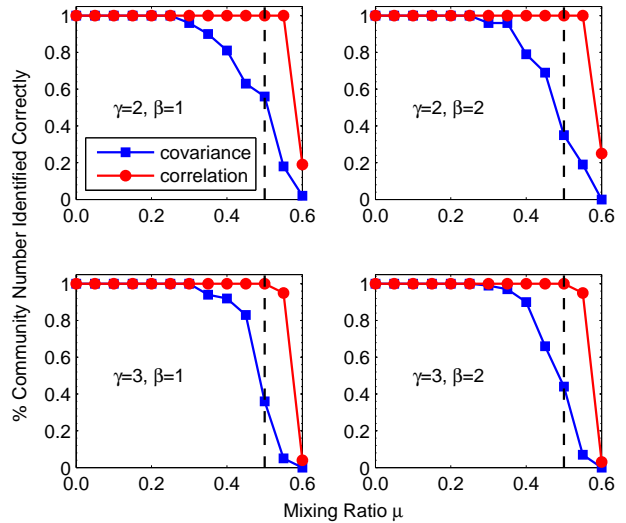}
\caption{\label{fig5}(Color online) Comparison between the
correlation matrix and the covariance matrix in terms of their
effectiveness at identifying the number of communities on benchmark
networks with different parameter configurations. For each parameter
configuration, $100$ generated networks are used.}
\end{figure}

\begin{figure}
\includegraphics[width=0.48\textwidth]{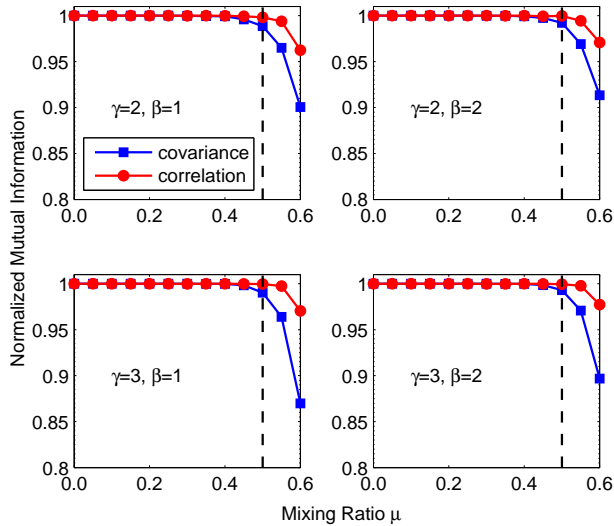}
\caption{\label{fig6}(Color online) Comparison between the
correlation matrix and the covariance matrix in terms of their
effectiveness at identifying the community structure on benchmark
networks with different parameter configurations. Each point
corresponds to an average over $100$ network realizations.}
\end{figure}

Note that each benchmark network has only one intrinsic topological
scale according to the construction rules. Thus we only consider the
largest eigengap in the spectrum of the covariance matrix and the
correlation matrix. The communities are identified using the top $p$
eigenvectors corresponding to the eigenvalues lying the greater side
of the largest eigengap. Specifically, the top $p$ eigenvectors are
projected into the node vectors according to Eq.~(\ref{eq8}), and
then the communities are identified through clustering these node
vectors using the $k$-means clustering method. Note that the
community number is $p+1$.

The first test focuses on whether the intrinsic scale can be
correctly uncovered. We investigate this problem through judging
whether the number of intrinsic communities can be correctly
identified. Fig.~\ref{fig5} shows the comparison between the
correlation matrix and the covariance matrix (identically the
modularity matrix) on networks with four different parameter
configurations for $\gamma$ and $\beta$. When the community
structure is evident, i.e., the mixing ratio $\mu$ is smaller, both
the covariance matrix and the correlation matrix are effective at
identifying the correct number of communities and thus the intrinsic
scale of the network. However, when the community structure becomes
fuzzier with the increase of the mixing ratio $\mu$, the performance
of the covariance matrix deteriorates while the correlation matrix
still achieves rather good results. Even when the mixing ratio $\mu$
is larger than $0.5$, the border beyond which communities are no
longer defined in the strong sense~\cite{Radicchi04}, the number of
communities can still be accurately identified by investigating the
spectrum of the correlation matrix.

The second test turns to whether the intrinsic community structure
can be identified. As demonstrated by the first test, the
correlation matrix outperforms the covariance matrix at finding the
correct number of communities. However, as to the second test, we
assume that the community number has been given \textit{a priori}
and then we compare the effectiveness of the eigenvectors of these
two matrices in terms of the NMI when comparing the answer
partitions with the ones obtained by the methods based on these two
matrices. As shown in Fig.~\ref{fig6}, these two matrices both
exhibit very good performance at identifying the intrinsic community
structure. By comparison, the correlation matrix outperforms the
covariance matrix for all used parameter configurations. This
indicates that the eigenvectors of the correlation matrix
characterize the spread characteristics of network nodes better than
that of the covariance matrix, especially when the community
structure is fuzzier.

\subsection{Tests on real world networks}

\begin{figure}
\hspace{-0pt}
\includegraphics[width=0.48\textwidth]{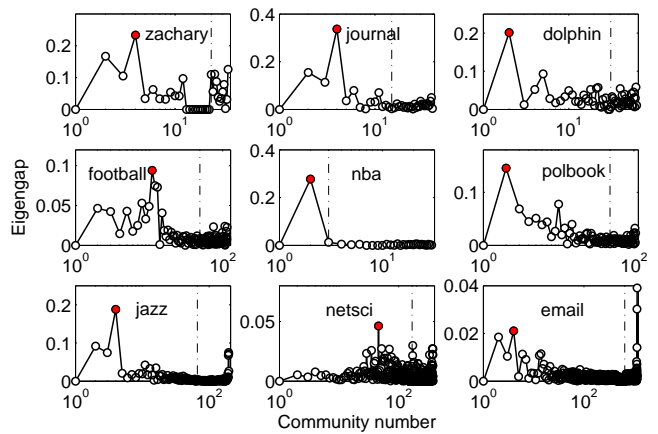}
\caption{\label{fig7}(Color online) Tests on real world networks.
The vertical axis represents the length of the eigengap and the
horizontal axis represents the the corresponding community number
indicated by the eigengap. The shaded-circles mark the largest
eigengap. Note that, in order to identify the community structure,
only the eigengaps between positive eigenvalues are taken into
account. The vertical dashed lines are the place where the
zero-valued eigenvalues occur.}
\end{figure}

In the previous subsection, we have demonstrated that the
correlation matrix is superior to the covariance matrix at
uncovering the intrinsic topological scale of the artificial
benchmark networks. Now we apply the method based on the correlation
matrix on real world networks. These networks are widely used to
evaluate community detection methods. These networks include the
Zachary's karate club network~\cite{Zachary77}, the journal index
network constructed in~\cite{Newman06a}, the social network of
dolphin by Lusseau \textit{et al}~\cite{Lusseau03}, the college
football network of the United States~\cite{Girvan02}, the match
network of the NBA teams, the network of political
books~\cite{Rosvall07}, the network of jazz
musician~\cite{Gleiser03}, the coauthor network of network scientist
presented in~\cite{Newman06b}, and the email network of
URV~\cite{Guimera03}. For convenience, these networks are
respectively abbreviated to \textit{Zachary}, \textit{journal},
\textit{dolphin}, \textit{football}, \textit{nba}, \textit{polbook},
\textit{jazz}, \textit{netsci} and \textit{email}. The test results
on these networks are shown in Fig.~\ref{fig7}.

As shown in Fig.~\ref{fig7}, the eigengaps of the correlation matrix
are very effective at the identification of the community number and
accordingly the intrinsic scale of the network. For most of these
networks with known community structure, the correlation matrix can
accurately uncover such community structure including the networks
\textit{journal}, \textit{dolphin}, \textit{nba} and
\textit{polbook}. For the other networks, the correlation matrix
also gives very promising results for their community structure.
Specifically, the detected communities reflect the structural and
functional characteristics of each specific network, which can be
verified through checking the nodes of each community. In addition,
for the last three networks, large eigengaps can be observed among
the negative eigenvalues. This indicates that these networks contain
anticommunity structure. Actually, this phenomenon is also observed
in many other real world networks, which are not included in this
article. This will be discussed in our further work.

In addition, we also test our method on some real world complex
networks with larger size, including the protein interaction network
and the word association network. However, no significant eigengap
is observed in the spectrum of the covariance matrix and the
correlation matrix associated with these networks. This indicates
that there is no scale which is significantly superior to other
scales and thus no partition of network is more desired. Different
from the eigenvalues of the covariance matrix, however, besides the
length of the eigengaps among eigenvalues, the magnitude itself of
the eigenvalues of the correlation matrix can provide us intuitive
knowledge on the cohesiveness of the communities at each specific
scale. Generally speaking, an eigenvalue larger than $0.5$ indicates
the existence of a cohesive node group, i.e., a community. Thus,
according to the magnitude of eigenvalues, people can choose the
topological scale or partition of network which is appropriate for
their application and practical demands.

\section{Conclusions and discussions}
\label{sec6}

In this article, we have studied the problem of detecting the
multi-scale community structure in networks from the perspective of
dimension reduction, i.e., from the straightforward description
taking each node as one dimension to a coarse-grained description
where each dimension corresponds to a community. The main
contributions of this article are as follows. Firstly, the
covariance matrix of network is defined to characterize the
relationship between original node dimensions through the
translation transformation. The covariance matrix is shown to be the
unbiased version of the traditional modularity matrix, which
underlies the benefit function of network partition, namely
modularity. Secondly, using the rotation transformation, we
recognize several reduced dimensions which are significant at
reflecting the community structure of network. Taking these reduced
dimensions as new axis vectors, the multi-scale community structure
is then identified through investigating the spectrum of the
covariance matrix and projecting the nodes of network into
low-dimensional node vectors. Thirdly, to deal with the
heterogeneous distributions of node degree and community size, the
correlation matrix is proposed through introducing the rescaling
transformation into the covariance matrix.

Extensive tests on real world and artificial networks demonstrate
that the correlation matrix significantly outperforms the covariance
matrix or the modularity matrix as regards revealing the multi-scale
community structure of network. Note that the computational efforts
of the correlation matrix are essentially identical to the
modularity matrix. Thus we suggest to use the correlation matrix and
accordingly the benefit function $F$ to detect the multi-scale
community structure of network, especially when the network has
heavily heterogeneous distribution of node degree and community
size. We also hope that the readers feel encouraged to apply the
correlation matrix to other real world networks and we look forward
to seeing the more efficient heuristic methods to optimize the
benefit function $F$.

In addition, we also notice that no significant scale is observed in
several real world complex networks. In this case, the covariance
matrix or the modularity matrix fails to provide obvious evident
about the community structure. However, since the magnitude itself
of the eigenvalues of the correlation matrix has indicative meaning
about the cohesiveness of communities, the correlation matrix can
still be used to reveal the community structure.

Finally, as to traditional spectral clustering methods, the
normalized Laplacian matrix is also been suggested to replace the
standard Laplacian matrix. The basic idea of the normalized
Laplacian matrix is consistent to the rescaling transformation
introduced in this article. Compared to the heuristic normalization
transformation on the standard Laplacian, the rescaling
transformation in this article has solid theoretical foundation from
the perspective of dimension reduction, for example, the principal
component analysis. Thus, the work in this article might provide
insights to understand the spectral clustering methods and might
have important implications to the clustering problem dealing with
datasets with heterogeneous cluster size.

\begin{acknowledgments}
This work was funded by the National Natural Science Foundation of
China under grant number $60873245$ and $60933005$. The authors
thank Alex Arenas, Mark Newman, Santo Fortunato, Martin Rosvall, and
David Lusseau for providing network and other data used in this
paper.
\end{acknowledgments}


\begin{thebibliography}{36}
\expandafter\ifx\csname url\endcsname\relax
  \def\url#1{\texttt{#1}}\fi
\expandafter\ifx\csname urlprefix\endcsname\relax\def\urlprefix{URL
}\fi

\bibitem{Girvan02}
M. Girvan and M.~E.~J. Newman, Community structure in social and
biological networks, \textit{Proc. Natl. Acad. Sci. U.S.A.}
\textbf{99}, 7821 (2002).

\bibitem{Newman03a}
M.~E.~J. Newman, The structure and function of complex networks,
\textit{SIAM Rev.} \textbf{45}, 167 (2003).

\bibitem{Flake02}
G.~W. Flake, S.~R. Lawrence, C.~L. Giles, and F.~M. Coetzee,
Self-organization and identification of Web communities,
\textit{IEEE Comput.} \textbf{35}, 66 (2002).

\bibitem{Cheng09}
X.~Q. Cheng, F.~X. Ren, S. Zhou and M.-B. Hu, Triangular clustering
in document networks, \textit{New J. Phys.} \textbf{11}, 033019
(2009).

\bibitem{Guimera05}
R.~Guimer\`a and L.~A.~N. Amaral, Functional cartography of complex
metabolic networks, \textit{Nature (London)} \textbf{433}, 895
(2005).

\bibitem{Newman04}
M.~E.~J. Newman and M. Girvan, Finding and evaluating community
structure in networks, \textit{Phys. Rev. E} \textbf{69}, 026113
(2004).

\bibitem{Newman06a}
M.~E.~J. Newman, Modularity and community structure in networks,
\textit{Proc. Natl. Acad. Sci. U.S.A.} \textbf{103}, 8577 (2006).

\bibitem{Newman06b}
M.~E.~J. Newman, Finding community structure in networks using the
eigenvectors of matrices, \textit{Phys. Rev. E} \textbf{74}, 036104
(2006).

\bibitem{Palla05}
G. Palla, I. Der\'enyi, I. Farkas, and T. Vicsek, Uncovering the
overlapping community structure of complex networks in nature and
society, \textit{Nature (London)} \textbf{435}, 814 (2005).

\bibitem{Radicchi04}
F. Radicchi, C. Castellano, F. Cecconi, V. Loreto, and D. Parisi,
Defining and identifying communities in networks, \textit{Proc.
Natl. Acad. Sci. U.S.A.} \textbf{101}, 2658 (2004).

\bibitem{Clauset04}
A. Clauset, M.~E.~J. Newman, and C. Moore, Finding community
structure in very large networks, \textit{Phys. Rev. E} \textbf{70},
066111 (2004).

\bibitem{Duch05}
J. Duch and A. Arenas, Community detection in complex networks using
extremal optimization, \textit{Phys. Rev. E} \textbf{72}, 027104
(2005).

\bibitem{Shen09a}
H.~W. Shen, X.~Q. Cheng. K. Cai, and M.-B. Hu, Detect overlapping
and hierarchical community structure in networks, \textit{Physica A}
\textbf{388}, 1706 (2009).

\bibitem{Shen09b}
H.~W. Shen, X.~Q. Cheng, and J.~F. Guo, Quantifying and identifying
the overlapping community structure in networks, \textit{J. Stat.
Mech.: Theory Exp.} P07042 (2009).

\bibitem{Bagrow08}
J. P. Bagrow, Evaluating local community methods in networks,
\textit{J. Stat. Mech.: Theory Exp.} P05001 (2008).

\bibitem{Carmi08}
S. Carmi, P. L. Krapivsky, and D. ben-Avraham, Partition of networks
into basins of attraction, \textit{Phys. Rev. E} \textbf{78},
066111, (2008).

\bibitem{Fortunato10}
S. Fortunato, Community detection in graphs, \textit{Phys. Rep.}
\textbf{486}, 75 (2010).

\bibitem{Fortunato07}
S. Fortunato and M. Barth\'{e}lemy, Resolution limit in community
detection, \textit{Proc. Natl. Acad. Sci. U.S.A.} \textbf{104}, 36
(2007).

\bibitem{Lambiotte08}
R. Lambiotte, J. C. Delvenne, and M. Barahona, Laplacian Dynamics
and Multiscale Modular Structure in Networks, \textit{e-print
arXiv}: 0812.1770, (2008).

\bibitem{Arenas08}
A. Arenas, A. Fern\'{a}ndez, and S. G\'{o}mez, Analysis of the
structure of complex networks at different resolution levels,
\textit{New J. Phys.} \textbf{10}, 053039 (2008).

\bibitem{Ronhovde09}
P. Ronhovde and Z. Nussinov, Multiresolution community detection for
megascale networks by information-based replica correlations,
\textit{Phys. Rev. E} \textbf{80}, 016109 (2009).

\bibitem{Arenas06}
A. Arenas, A. D\'{i}az-Guilera, and C. J. P\'{e}rez-Vicente,
Synchronization reveals topological scales in complex networks,
\textit{Phys. Rev. Lett.} \textbf{96}, 114102 (2006).

\bibitem{Cheng10}
X. Q. Cheng and H. W. Shen, Uncovering the community structure
associated with the diffusion dynamics on networks, \textit{J. Stat.
Mech.: Theory Exp.} P04024 (2010) .

\bibitem{Mucha10}
P. J. Mucha, T. Richardson, K. Macon, M. A. Porter, and J.-P.
Onnela, Community structure in time-dependent, multiscale, and and
multiplex networks, \textit{Science} \textbf{328}, 876, (2010).

\bibitem{Arenas10}
A. Arenas, J. Borge-Holthoefer, S. Gomez, and G. Zamora-Lopez,
Optimal map of the modular structure of complex networks,
\textit{New J. Phys.} \textbf{12}, 053009  (2010).

\bibitem{Fiedler73}
M. Fiedler, Praha: Algebraic connectivity of graphs, \textit{Czech.
Math. J.} \textbf{23}, 298 (1973).

\bibitem{Newman03b}
M.~E.~J. Newman, Mixing patterns in networks, \textit{Phys. Rev. E}
\textbf{67}, 026126 (2003).

\bibitem{Newman02b}
M.~E.~J. Newman, Assortative mixing in networks, \textit{Phys. Rev.
Lett.} \textbf{89}, 208701 (2002).

\bibitem{Zachary77}
W. W. Zachary, An information flow model for conflict and fission in
small groups, \textit{J. Anthropol. Res.} \textbf{33}, 452 (1977).

\bibitem{Chauhan09}
S. Chauhan, M. Girvan, and E. Ott, Spectral properties of networks
with community structure, \textit{Phys. Rev. E} \textbf{80}, 056114
(2009).

\bibitem{Eriksen03}
K. A. Eriksen, I. Simonsen, S. Maslov, and K. Sneppen. Modularity
and extreme edges of the internet, \textit{Phys. Rev. Lett.}
\textbf{90}, 148701 (2003).

\bibitem{Capocci05}
C. Capocci, V. D. P. Servedio, G. Caldarelli, and F. Colaiori,
Detecting communities in large networks, \textit{Physica A}
\textbf{352}, 669 (2005).

\bibitem{Lancichinetti08}
A. Lancichinetti, S. Fortunato, and F. Radicchi, Benchmark graphs
for testing community detection algorithms, \textit{Phys. Rev. E}
\textbf{78}, 046110 (2008).

\bibitem{Danon2005}
L.~Danon, J.~Duch, A.~D\`{i}az-Guilera, J.~Duch and A.~Arenas,
Comparing community structure identification, \textit{J. Stat.
Mech.: Theory Exp.}, P09008 (2005).

\bibitem{Lusseau03}
D. Lusseau, K. Schneider, O. J. Boisseau, P. Haase, E. Slooten, and
S. M. Dawson, The bottlenose dolphin community of Doubtful Sound
features a large proportion of long-lasting associations: Can
geographic isolation explain this unique trait? \textit{Behav. Ecol.
Sociobiol.} \textbf{54} 396 (2003).

\bibitem{Rosvall07}
M. Rosvall and C. T. Bergstrom, An information-theoretic framework
for resolving community structure in complex networks, \textit{Proc.
Natl. Acad. Sci. U.S.A.} \textbf{104}, 7327 (2007).

\bibitem{Gleiser03}
P. Gleiser and L. Danon, Community structure in jazz, \textit{Adv.
Complex Syst.} \textbf{6}, 565 (2003).

\bibitem{Guimera03}
R. Guimer\`{a}, L. Danon, A. D\'{i}az-Guilera, F. Giralt, and A.
Arenas, Self-similar community structure in a network of human
interactions, \textit{Phys. Rev. E} \textbf{68}, 065103(R) (2003).

\end{thebibliography}
\end{document}